\documentstyle[prl,aps,multicol,epsf]{revtex}
\def\narrowcaption{\columnwidth20.5pc}
\begin{document}
\draft
\author{Eugene M. Chudnovsky$^1$ and Jonathan R. Friedman$^2$}
\address{$^1$ Department of Physics and Astronomy, CUNY Lehman College\\
250 Bedford Park Boulevard West, Bronx, NY 10468-1589} \address{$^2$
Department of Physics and Astronomy, SUNY at Stony Brook\\ Stony Brook, NY
11794-3800}
\title{Macroscopic Quantum Coherence in a Magnetic Nanoparticle Above the
Surface of a Superconductor}
\date{\today}
\maketitle

\begin{abstract}
We study macroscopic quantum tunneling of the magnetic moment in a
single-domain particle placed above the surface of a superconductor. 
Such a setup allows one to manipulate the height of the energy barrier,
preserving the degeneracy of the ground state. The tunneling amplitude and
the effect of the dissipation in the superconductor are computed.  
\end{abstract}


\smallskip

\begin{multicols}{2}
Tunneling of the magnetic moment in nanoparticles and molecular clusters
has been intensively studied theoretically and experimentally in the last
decade \cite{book}. The interest in this problem is two-fold. Firstly,
magnetic 
tunneling reveals itself at a quasiclassical level, that is, in situations
where 
all three components of the magnetic moment, ${\bf M}$, can be rather
accurately 
determined by a macroscopic measurement. The interaction of ${\bf M}$ with
microscopic degrees of freedom makes this problem one of tunneling with
dissipation \cite{CL}. Secondly, tunneling of the magnetic moment changes
the magnetic properties of small magnets, with potential implications for
the data- storage technology. It also adds nanomagnets to the list of
candidates for qubits - the elements of quantum computers.

In zero magnetic field the magnetic state of a classical magnet is
degenerate with respect to ${\bf M}{\rightarrow}-{\bf M}$ due to
time-reversal symmetry. 
In 
nanoparticles the $|\uparrow>$ and $|\downarrow>$ minima of the energy are
separated by a barrier, $U$, due to the magnetic anisotropy. The thermal
rate 
of switching between the two classical states is proportional to
$\exp(-U/T)$. 
At high temperatures, when the thermal rate is high, the particle is in the
superparamagnetic regime. At low temperature, as far as the thermal rate is
concerned, the magnetic moment should freeze along one of the anisotropy
directions. In particles of sufficiently small size, however, even at
$T=0$ the magnetic moment can switch due to quantum tunneling. The quantum
switching rate, ${\Gamma}$, scales with the total spin $S$ of the
nanoparticle according \cite{EMC,Schilling,Hemmen,CG,Garanin} to
$\ln{\Gamma}{\propto}-S$. If the switching time, ${\Gamma}^{-1}$, is small
compared to the measuring time, the nanoparticle remains superparamagnetic
in the limit of $T=0$. If, in addition to 
that, the interaction of ${\bf M}$ with microscopic degrees of freedom
(phonons, 
itinerant electrons, nuclear spins, etc.) is small, the nanoparticle,
during a certain decoherence time, can exist in a coherent quantum
superposition of two classical states, ${\Psi}_o=|\uparrow>+|\downarrow>$.
In that state the probability that the moment of the particle has a certain
 orientation oscillates in time as $\cos({2\Gamma}t)$. 

For a magnetic particle of considerable size to be in the quantum
superparamagnetic (not necessarily coherent) regime, the energy barrier
between 
the $|\uparrow>$ and $|\downarrow>$ states must be made sufficiently small. 
One 
way to decrease the barrier is to use an external magnetic field. This
method has been used in experiments on magnetic particles and molecular
clusters performed to date \cite{Awschalom,Friedman,Wernsdorfer}. It has a
clear drawback 
if one attempts to create a coherent superposition of the $|\uparrow>$ and
$|\downarrow>$ states. Namely, the external field, unless it is applied
exactly 
parpendicular to the anisotropy axis, removes the degeneracy between the
$|\uparrow>$ and $|\downarrow>$ states. In this Letter, we suggest a method
of controlling the barrier without breaking the degeneracy. This method is
illustrated in Fig. 1. The nanoparticle is placed above the surface of a
superconductor at a variable distance controlled by, e.g., a piezoelectric
layer 
or holder. The current induced in the superconductor creates the magnetic
image 
of the nanoparticle. The interaction between the nanoparticle and the
superconductor is then equivalent to the dipole interaction between the
nanoparticle and its image. The reduction of the barrier is similar to the
one from the external magnetic field applied opposite to ${\bf M}$.
However, contrary to the situation with the external field, the system (the
nanoparticle 
plus the superconductor) is now degenerate with respect to ${\bf
M}{\rightarrow}-
{\bf M}$. It should be emphasized that such a degeneracy is a very general
property of the system that is independent of the shape of the particle 
and the landscape of the superconducting surface. It is rooted in the
time-reversal symmetry of the system in the absence of the field. 
The tunneling rate for
the situation shown in Fig. 1 will be computed below. 

A high tunneling rate does not automatically provide the coherent
superposition of quasiclassical states. Different mechanisms of decoherence
due 
to interactions inside the nanoparticle have been worked out in recent
years \cite{Garg,Stamp,Levine}. At low temperature, in the absence of
nuclear spins and itinerant electrons, the effect of dissipation on the
tunneling rate can be very 
small \cite{book}. Decoherence is a more subtle issue. Generally speaking,
macroscopic quantum coherence (MQC), that is $\cos({2\Gamma}t)$ oscillations
of ${\bf M}$, occur only if the decohering interactions are small compared
to the tunnel splitting of the ground state, ${\hbar}{\Gamma}$. We will
show that this condition can be satisfied at least as far as the
interaction between the nanoparticle and the superconductor is concerned.
This question is of interest also in a more general context of measuring
the rotation of a mesoscopic spin with the help of a superconducting
device. Indeed, most of the experiments on individual nanoparticles used
SQUIDs. Although the problem studied here is different from those
experimental situations, some of the ideas should also apply in those cases. 
 
Let us consider tunneling in a nanoparticle above the flat surface of a
superconductor \cite{note}, as shown in Fig. 1, in the absence of dissipation. 
The external magnetic field will be considered zero throughout this paper.
To make the classical electrodynamics of the problem less cumbersome, we
will make 
certain simplifying assumptions about the superconductor, the shape of the
particle and its magnetic anisotropy. None of them is important and any
generalization can be studied along the same lines. First, we shall assume
that 
the superconductor is in the Meissner regime; that is, the magnetic field
at the 
surface of the superconductor does not exceed $H_{c1}$. We shall also
assume that the characteristic 
geometrical dimensions of the problem, the size of the particle and its
distance 
to the surface, are large compared to the penetration depth ${\lambda}_{L}$. 
In 
that case, it is a good approximation to take the field of the particle at
the surface of the superconductor to be parallel to that surface. The
effect of the 
superconductor on the particle is then equivalent to the magnetic dipole
interaction with the image shown in Fig. 1. 

Next we assume that the particle is of ellipsoidal shape (that is,
uniformly 
magnetized) with crystal fields either small or dominated by the single-ion
anisotropy. The total energy of the magnetic anisotropy of such a particle
must 
be quadratic in the magnetization \cite{LL}, \begin{equation}
E_{an}=\frac{2{\pi}}{V}N_{ik}M_{i}M_{k}\;\;\;, \end{equation}
where ${\bf M}$ is the total magnetic moment of the particle, $V$ is its
volume, 
and the tensor $N_{ik}$ includes both the demagnetizing effect (that is,
shape anisotropy) and the magnetocrystalline anisotropy. In a ferromagnetic
particle, 
$M$ is proportional to $V$, while $N_{ik}$ is independent of $V$. The
factor $V^{-1}$ in Eq. (1) is, therefore, needed to provide the correct
linear scaling of $E_{an}$ with $V$. We shall assume that the principal
axes of $N_{ik}$ coincide with the coordinate axes in Fig. 1; with $Z$ being
the easy magnetization direction and $N_{xx}>N_{yy}>N_{zz}$. The magnitude of
${\bf M}$ is assumed, as usual, to be formed by a strong exchange
interaction and, thus, independent of the orientation.  That is,
$M_{x}^{2}+M_{y}^{2}+M_{z}^{2}=M^{2}=const.$ The energy of the magnetic
anisotropy then becomes \begin{equation}
E_{an}=\frac{1}{V}({\beta}_{x}M_{x}^{2}-{\beta}_{z}M_{z}^{2})\;\;\;,
\end{equation}
where ${\beta}_{x},{\beta}_{z}$ are positive dimensionless coefficients of
order unity.

Eq. (2) describes a magnet having a $YZ$ easy magnetization plane with $Z$
being 
the easy axis in that plane. The two degenerate minima of Eq. (2)
correspond to 
${\bf M}$ looking along and opposite to the $Z$ axis. In the Meissner state
of the superconductor, the magnetic field of the particle induces
superconducting currents whose field is equivalent to the field of the
image shown in Fig. 1. 
As 
the particle moves closer to the superconductor, the interaction between
the particle and its image increases and the barrier between the two
equilibrium orientations of ${\bf M}$ decreases. The magnetic moment of the
particle, ${\bf 
M}$, and the moment of the image,  ${\bf m}$, are related through
\begin{equation}
M_{x}=m_{x}\,,\;\;M_{y}=m_{y}\,,\;\;M_{z}=-m_{z}\;\;\;.
\end{equation}
With reasonable accuracy the energy of the magnetic dipole interaction
between the particle and its image is given by
\begin{equation}
E_{int}=\frac{[{\bf M}{\cdot}{\bf m}-3({\bf n}{\cdot}{\bf M}) ({\bf
n}{\cdot}{\bf m})]}{(2d)^{3}}\;\;\;, \end{equation}
where $d$ is the distance from the (center of the) particle to the surface
of the superconductor. With the help of relations (4), one
obtains\cite{note} \begin{equation}
E_{int}=\frac{M_{z}^{2}}{(2d)^{3}}\;\;\;.
\end{equation}
The total energy of the system, $E=E_{an}+E_{int}$ then becomes
\begin{equation}
E=\frac{1}{V}(-{\beta}_{z}{\epsilon}M_{z}^{2}+{\beta}_{x}M_{x}^{2})\;\;\;,
\end{equation}
where we have introduced ${\epsilon}=1-V/{\beta}_{z}(2d)^{3}$. 

According to Eq. (6) the energy barrier between the degenerate 
${|\uparrow>}$
and ${|\downarrow>}$ states of the particle is given by
$U={\beta}_{z}{\epsilon}M_{o}^{2}V$, where $M_{o}=M/V$ is the volume 
magnetization of the particle. Our main idea is to manipulate $d$ in
such a way that ${\epsilon}$ and, consequently, $U$ become small enough to
provide a significant tunneling rate. Note that most deviations from the
simplifying assumptions made above will renormalize ${\beta}_{x,z}$ and 
$d$ in Eq. (6) but will not change the form of the total energy. It is
convenient 
to introduce the total dimensionless spin of the particle, ${\bf S}={\bf
M}/{\hbar}{\gamma}$ (${\gamma}$ being the gyromagnetic ratio) and 
two characteristic frequencies:
\begin{equation}
{\omega}_{\parallel}={\beta}_{z}{\epsilon}{\gamma}M_{o}\;,
\;\;\;{\omega}_{\perp}={\beta}_{x}{\gamma}M_{o}\;\;.
\end{equation} 
The total energy can be then written as \begin{equation}
E=\frac{1}{S}[-
{\hbar}{\omega}_{\parallel}S_{z}^{2}+{\hbar}{\omega}_{\perp}S_{x}^{2}]\;\;\;.
\end{equation}
The corresponding tunneling problem has been studied by a number of authors
\cite{Schilling,Hemmen,CG,Garanin}. In the limit of
${\omega}_{\parallel}<<{\omega}_{\perp}$, i.e. at ${\epsilon}<<1$, the
tunneling rate at $T=0$ in the absence of dissipation is given by
${\Gamma}_{o}=A_{o}\exp(-B_{o})$ with
\begin{eqnarray}
A_{o} & = & 
\frac{16}{\sqrt{\pi}}S^{1/2}{\omega}_{\parallel}^{3/4}{\omega}_{\perp}^{1/4}
 \nonumber \\
B_{o} & = & 
2S\left(\frac{{\omega}_{\parallel}}{{\omega}_{\perp}}\right)^{1/2}\;\;\;.
\end{eqnarray}

So far, we have neglected non-dissipative terms in the total energy that
come from the superconductor. One such term is the kinetic energy of the
Cooper pairs,
\begin{equation}
E_{sc1}=\int\,d^{3}r\frac{n_{s}m{\bf v}_{s}^{2}}{2}\;\;\;, \end{equation}
where $n_{s}$ is the concentration of superconducting electrons, $m$ is
their mass, and ${\bf v}_{s}={\bf j}_{s}/en_{s}$ is their drift velocity
expressed in 
terms of the superconducting current $j_{s}$. Eq. (10) can be written as
\begin{equation}
E_{sc1}=\frac{4{\pi}{\lambda}_{L}^{2}}{c^{2}}\int\,d^{3}r{\bf
j}_{s}^{2}\;\;\;, \end{equation}
where ${\lambda}_{L}=mc^{2}/4{\pi}e^{2}n_{s}$ is the London penetration depth.
The superconducting current is concentrated near the surface, resulting in
the surface current \cite{LL}
\begin{equation}
{\bf g}_{s}=\int\,dz{\bf j}_{s}=\frac{c}{4{\pi}}{\bf n}{\times}{\bf B}({\bf
r})\;\;\;,
\end{equation}
where ${\bf n}$ is the unit vector in the $Z$ direction and ${\bf B}({\bf
r})$ is the sum of the dipole fields of the magnetic particle and its image
at $z=0$.
A somewhat tedious but straightforward calculation then gives \begin{equation}
E_{sc1}=\frac{{\lambda}_{L}}{8{\pi}}\int\,d^{2}r{\bf B}^{2}({\bf r})=const
+ \frac{3{\lambda}_{L}}{16d^{4}}M_{z}^{2}\;\;\;.
\end{equation}
The contribution of the kinetic energy of the superconducting electrons to
the total energy is small if $d>>{\lambda}_{L}$. Even at
$d{\sim}{\lambda}_{L}$, however, it reduces to the renormalization of the
$d$ dependence of the interaction between the magnetic particle and 
the superconductor. Since we are interested in $d$ close 
to the critical value at which the barrier becomes zero, the form of the
Hamiltonian, Eq. (8), and the expressions, Eq. (9), for the tunneling rate
remain unaffected by that renormalization.

The next non-dissipative term in the energy is the inertia coming from the
energy of the electric field,
\begin{equation}
E_{sc2}=\int\,d^{3}r\frac{{\bf E}^{2}}{8{\pi}}=
\frac{1}{8{\pi}c^{2}}\int\,d^{3}r\left(\frac{d{\bf A}}{dt}\right)^{2}\;\;\;.
\end{equation}
where ${\bf A}=-(4{\pi}{\lambda}_{L}^{2}/c){\bf j}_{s}$ is the vector
potential 
in the superconductor. Then, similarly to the previous case, one obtains
\begin{equation}
E_{sc2}=\frac{{\lambda}_{L}^{3}}{16{\pi}c^{2}}\int\,d^{2}r \left(\frac{d{\bf 
B}}{dt}\right)^{2}=\frac{3{\lambda}_{L}^{3}}{32c^{2}d^{4}}({\dot{\bf
M}}^{2}+ {\dot{M}}_{z}^{2})\;\;\;.
\end{equation}
To estimate the effect of this inertia term on tunneling, it should be
compared, at the instanton frequency\cite{book}
${\omega}_{inst}=({\omega}_{\parallel}{\omega}_{\perp})^{1/2}$, with $E$ of
Eq. (8). The ratio of the energies is \begin{equation}
E_{sc2}/E\;{\sim}\;({\lambda}_{L}/d)^{3}(l/d)(l/{\lambda}_{inst})^{2}\;\;\;,
 \end{equation}
where we have introduced $l=V^{1/3}$ and ${\lambda}_{inst}=c/{\omega}_{inst}$.
Even if the experimental values of $l$,$d$ and ${\lambda}_{L}$ do not
differ in 
order of magnitude, ${\lambda}_{inst}$ can hardly be less than 1 cm, making
the 
ratio of energies in Eq. (16) negligible for nanoparticles used in tunneling
experiments. We may then conclude that Eq. (9) gives a good estimate of the
tunneling rate in the absence of dissipation.

Spin tunneling with dissipation due to the interaction of ${\bf M}$ with
microscopic 
degrees of freedom inside the nanoparticle has been intensively studied \cite{Garg,Stamp}. Interactions with phonons, magnons, nuclear
spins, etc. 
have been considered. Here we will study the mechanism of dissipation
specific to our problem: the interaction of the magnetic moment with normal
quasiparticles in the superconductor. This analysis may also be relevant to
experiments on spin tunneling performed using SQUIDs. 

We begin with the derivation of the energy dissipation in the
superconductor due 
to the rotation of ${\bf M}$. With the help of the relations ${\bf E}=-c^{-
1}{\partial}{\bf A}/{\partial}t$ and 
${\bf j}={\sigma}{\bf E}$, one obtains
\begin{equation}
Q=\int\,d^{3}r{\bf j}{\cdot}{\bf E}=\frac{1}{c^{2}}\int\,d^{3}r
{\sigma}(t){\dot{\bf A}}^{2}\;\;\;.
\end{equation}
where ${\sigma}$ is the conductivity due to quasiparticles. Its Fourier
transform can be approximated as ${\sigma}_{\omega}=e^{2}n_{q}/m({\nu}+
i{\omega})$, where $n_{q}$ is the quasiparticle concentration and
${\nu}$ is their scattering rate. The latter is typically 2-3 orders of
magnitude greater than the instanton frequency for the tunneling of ${\bf M}$.
Consequently, the time dependence of ${\sigma}$ in Eq. (17) can be neglected
and one can use ${\sigma}=e^{2}n_{q}(T)/m{\nu}$. Equation (17) then becomes
similar to Eq. (14) and the same argument gives the surface integral that
is proportional to $({\dot{\bf M}}^{2}+ {\dot{M}}_{z}^{2})\propto
[{\dot{\theta}}^{2}+({\dot{\theta}}^{2}+{\dot{\phi}}^{2}){\sin}^{2}{\theta}]$.
Due to the smallness of $\epsilon$, the hard-axis anisotropy
in Eq. (8) is small compared to the easy-axis anisotropy. Under this condition,
quasiclassical trajectories of ${\bf M}$ must be close to the easy plane. This
means ${\phi}{\approx}{\pi}/2$, while ${\theta}$ for the tunneling
trajectory changes from $0$ to $\pi$. A more rigorous analysis shows that 
${\dot{\phi}}^{2}{\sim}{\epsilon}{\dot{\theta}}^{2}$. Thus, with good
accuracy,
\begin{equation}
Q=\frac{{\sigma}{\lambda}_{L}^{3}}{2c^{2}}\int\,d^{2}r{\dot{\bf B}}^{2}=
\frac{3{\lambda}_{L}M^{2}}{16{\nu}d^{4}}\left(\frac{n_{q}}{n_{s}}\right)
{\dot{\theta}}^{2}(1+{\sin}^{2}{\theta})\;\;\;.
\end{equation}
If Eq. (18) were quadratic in ${\dot{\theta}}$, it could be interpreted as
linear dissipation with a friction coefficient
${\eta}=3{\lambda}_{L}M^{2}n_{q}/16{\nu}d^{4}n_{s}$. This would allow the
Caldeira-Leggett \cite{CL} approach to tunneling with dissipation. 
The dissipation in the rotation of ${\bf M}$ due to its interaction with
quasiparticles is nonlinear in ${\theta}$, however. Nevertheless, since 
we only want to obtain an estimate of the effect of dissipation on
tunneling, and because the ${\sin}^{2}{\theta}$ term in Eq. (18) can hardly
change this effect significantly, we shall go ahead and estimate the
effective Caldeira-Leggett action as 
\begin{equation}
I_{CL}=\frac{\eta}{4{\pi}}\int_{\infty}^{\infty}d{\tau}'\int_{o}^{\infty}
d{\tau}\frac{[{\theta}(\tau)-{\theta}({\tau}')]^{2}}{({\tau}-{\tau}')^{2}}\;
\;.
\end{equation}
Note that the double-integral in Eq. (19) is dimensionless. 

The measure of the effect of dissipation on tunneling is the ratio
$I_{CL}/{\hbar}B_{o}$ where ${\hbar}B_{o}$ is the effective action in the absence
of dissipation, with $B_{o}$ given by Eq. (9). After simple algebra, one
obtains \begin{equation}
\frac{I_{CL}}{{\hbar}B}{\sim}\frac{3{\pi}}{128{\sqrt{\epsilon}}}
\left(\frac{n_{q}}{n_{s}}\right)
\left(\frac{{\lambda}_{L}}{d}\right)\left(\frac{l}{d}\right)^{3}
\left(\frac{{\gamma}M_{o}}{\nu}\right)\;\;\;.
\end{equation}
In a typical experiment, one should expect ${\lambda}_{L}<d$ and $l<d$,
while the ratio ${\gamma}M_{o}/{\nu}$ may hardly exceed $10^{-2}$.
Consequently, even 
at $T{\sim}T_{c}$, when $n_{q}{\sim}n_{s}$, the effect of the dissipation
on the 
tunneling rate may become visible only at ${\epsilon}$ less than $10^{-5}$. 

A more rigorous approach along the lines of Ref. \cite{Tinkham} shows that
the ratio $n_{q}/n_{s}$ in equations (19)-(21) should be replaced by the
ratio of the coherence factors. Similar to other dissipation problems due
to quasiparticles \cite{Tinkham}, its effect on tunneling must have a
maximum at $T$ slightly lower than $T_{c}$. Even at the maximum this effect
should still be small. 
Coherence is a more subtle issue. It may be destroyed by a dissipative
environment even if the effect of dissipation on tunneling is weak. The
mechanism of dissipation discussed above goes down with temperature as
$\exp(- {\Delta}/T)$, where ${\Delta}$ is the superconducting gap.
Consequently, preserving coherent oscillations of ${\bf M}$ between
${|\uparrow>}$ and ${|\downarrow>}$ states requires $T<<{\Delta}$. 

The work of E.M.C. has been supported by the U.S. Department of Energy
through Grant No. DE-FG02-93ER45487.

\begin{figure}[p]
\narrowcaption
\caption{Magnetic particle above the surface of a superconductor.
The effect of the superconductor on the particle is equivalent to the field
from the image shown in the figure.} \end{figure}  

\end{multicols}


\begin{references}
\bibitem{note} In general, the surface of the
superconductor need not be flat. 
Other cases (e.g. when the superconductor is in the shape of a
hemispherical mirror) may give a relatively larger interaction energy 
than Eq. (5) and, hence, may be more experimentally relevant.
\end{references}
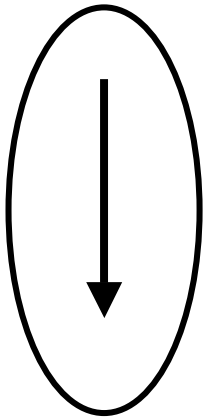
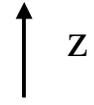
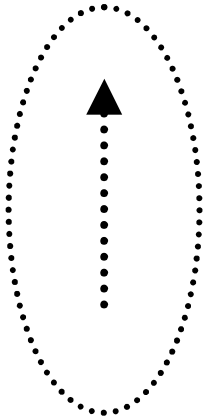
\end{document}